# Image Restoration Using Joint Statistical Modeling in Space-Transform Domain

Jian Zhang, Debin Zhao, Ruiqin Xiong, Siwei Ma, Wen Gao, *Fellow, IEEE*

*Abstract*—This paper presents a novel strategy for high-fidelity image restoration by characterizing both local smoothness and nonlocal self-similarity of natural images in a unified statistical manner. The main contributions are three-folds. First, from the perspective of image statistics, a joint statistical modeling (JSM) in an adaptive hybrid space-transform domain is established, which offers a powerful mechanism of combining local smoothness and nonlocal self-similarity simultaneously to ensure a more reliable and robust estimation. Second, a new form of minimization functional for solving image inverse problem is formulated using JSM under regularization-based framework. Finally, in order to make JSM tractable and robust, a new Split-Bregman based algorithm is developed to efficiently solve the above severely underdetermined inverse problem associated with theoretical proof of convergence. Extensive experiments on image inpainting, image deblurring and mixed Gaussian plus salt-and-pepper noise removal applications verify the effectiveness of the proposed algorithm.

*Index Terms*—Image restoration, statistical modeling, optimization, image inpainting, image deblurring

## I. Introduction

As a fundamental problem in the field of image processing, image restoration has been extensively studied in the past two decades [1]–[12]. It aims to reconstruct the original high quality image $x$ from its degraded observed version $y$, which is a typical ill-posed linear inverse problem and can be generally formulated as:

$$y = Hx + n, \qquad (1)$$

where $x, y$ are lexicographically stacked representations of the original image and the degraded image, respectively, $H$ is a matrix representing a non-invertible linear degradation operator and $n$ is usually additive Gaussian white noise. When $H$ is identity, the problem becomes image denoising [4], [5], [11]; when $H$ is a blur operator, the problem becomes image deblurring [14], [21]; when $H$ is a mask, that is, $H$ is a diagonal matrix whose diagonal entries are either 1 or 0, keeping or killing the corresponding pixels, the problem becomes image inpainting [22], [35]; when $H$ is a set of random projections, the problem becomes compressive sensing [16], [17]. In this paper, we focus on image inpainting, image deblurring and image denoising.

In order to cope with the ill-posed nature of image restoration, one type of scheme in literature employs image prior knowledge for regularizing the solution to the following minimization problem [14], [15]:

$$\operatorname{argmin}_x \tfrac{1}{2}\|Hx - y\|_2^2 + \lambda \Psi(x), \qquad (2)$$

where $\tfrac{1}{2}\|Hx - y\|_2^2$ is the $\ell_2$ data-fidelity term, $\Psi(x)$ is called the regularization term denoting image prior and $\lambda$ is the regularization parameter. In fact, the above regularization-based framework (2) can be strictly derived from Bayesian inference with some image prior possibility model. Many optimization approaches for regularization-based image inverse problems have been developed [13]–[15], [41], [42].

It has been widely recognized that image prior knowledge plays a critical role in the performance of image restoration algorithms. Therefore, designing effective regularization terms to reflect the image priors is at the core of image restoration.

Classical regularization terms utilize local structural patterns and are built on the assumption that images are locally smooth except at edges. Some representative works in the literature are total variation (TV) model [2], [14], half quadrature formulation [18] and Mumford-Shah (MS) model [20]. These regularization terms demonstrate high effectiveness in preserving edges and recovering smooth regions. However, they usually smear out image details and cannot deal well with fine structures, since they only exploit local statistics, neglecting nonlocal statistics of images.

In recent years, perhaps the most significant nonlocal statistics in image processing is nonlocal self-similarity exhibited by natural images. The nonlocal self-similarity depicts the repetitiveness of higher level patterns (e. g., textures and structures) globally positioned in images, which is first utilized to synthesize textures and fill in holes in images [19]. The basic idea behind texture synthesis is to determine the value of the hole using similar image patches, which also influences the image denoising task. Buades *et al.* [24] generalized this idea and proposed an efficient denoising model called nonlocal means (NLM), which takes advantage of this image property to conduct a type of weighted filtering for denoising tasks by means of the degree of similarity among surrounding pixels. This simple weighted approach is quite effective in generating sharper image edges and preserving more image details.

Later, inspired by the success of nonlocal means (NLM) denoising filter, a series of nonlocal regularization terms for

Manuscript received April 17, 2013; revised July 20, 2013; accepted November 7, 2013. This work is supported in part by the Major State Basic Research Development Program of China (2009CB320905) and National Science Foundation (No. 61272386, 61370114, and 61103088).

J. Zhang and D. Zhao are with the School of Computer Science and Technology, Harbin Institute of Technology, Harbin 150001, China (e-mail: jzhangcs@hit.edu.cn; dbzhao@hit.edu.cn).
R. Xiong, S. Ma, and W. Gao are with the National Engineering Laboratory for Video Technology, and Key Laboratory of Machine Perception (MoE), School of Electrical Engineering and Computer Science, Peking University, Beijing 100871, China (e-mail: rqxiong@pku.edu.cn; swma@pku.edu.cn; wgao@pku.edu.cn).

inverse problems exploiting nonlocal self-similarity property of natural images are emerging [25]–[29]. Note that the NLM-based regularizations in [25] and [28] are conducted at pixel level, i.e., from one pixel to another pixel. In [9] and [39], block-level NLM based regularization terms were introduced to address image deblurring and super-resolution problems. Gilboa and Osher defined a variational framework based on nonlocal operators and proposed nonlocal total variation (NL/TV) model in [25]. The connection between the filtering methods and spectral bases of the nonlocal graph Laplacian operator were discussed by Peyré in [27]. Recently, Jung *et al.* [29] extended traditional local MS regularizer and proposed a nonlocal version of the approximation of MS regularizer (NL/MS) for color image restoration, such as deblurring in the presence of Gaussian or impulse noise, inpainting, super-resolution, and image demosaicking.

Due to the utilization of self-similarity prior by adaptive nonlocal graph, nonlocal regularization terms produce superior results over the local ones, with sharper image edges and more image details [27]. Nonetheless, there are still plenty of image details and structures that cannot be recovered accurately. The reason is that the above nonlocal regularization terms depend on the weighted graph, while it is inevitable that the weighted manner gives rise to disturbance and inaccuracy [28]. Accordingly, seeking a method which can characterize image self-similarity powerfully is one of the most significant challenges in the field of image processing.

Based on the studies of previous work, two shortcomings have been discovered. On one hand, only one image property used in regularization-based framework is not enough to obtain satisfying restoration results. On the other hand, the image property of nonlocal self-similarity should be characterized by a more powerful manner, rather than by the traditional weighted graph. In this paper, we propose a novel strategy for high-fidelity image restoration by characterizing both local smoothness and nonlocal self-similarity of natural images in a unified statistical manner. Part of our previous work has been published in [30]. Our main contributions are listed as follows. First, from the perspective of image statistics, we establish a joint statistical modeling (JSM) in an adaptive hybrid space and transform domain, which offers a powerful mechanism of combining local smoothness and nonlocal self-similarity simultaneously to ensure a more reliable and robust estimation. Second, a new form of minimization functional for solving image inverse problems is formulated using JSM under regularization-based framework. The proposed method is a general model that includes many related models as special cases. Third, in order to make JSM tractable and robust, a new Split-Bregman based algorithm is developed to efficiently solve the above severely underdetermined inverse problem associated with theoretical proof of convergence.

The remainder of the paper is organized as follows. Section II elaborates the design of joint statistical modeling. Section III proposes a new objective functional containing a data-fidelity term and a regularization term formed by JSM, and gives the implementation details of solving optimization. Extensive experimental results are reported in Section IV. In Section V, we summarize this paper.

## II. PROPOSED JOINT STATISTICAL MODELING IN SPACE-TRANSFORM DOMAIN

As mentioned in Section I, to cope with the ill-posed nature of image inverse problems, the prior knowledge about natural images is usually employed, namely, image properties, which essentially play a key role to achieve high-quality images.

Here, two types of popular image properties are considered, namely local smoothness and nonlocal self-similarity, as illustrated by image Lena in Fig. 1. The former type describes the piecewise smoothness within local region, as shown by circular regions, while the latter one depicts the repetitiveness of the textures or structures in natural images globally positioned image patches, as shown by block regions with the same color. The challenge is how to characterize and formulate these two image properties mathematically. Note that different formulations of these two properties will lead to different results.

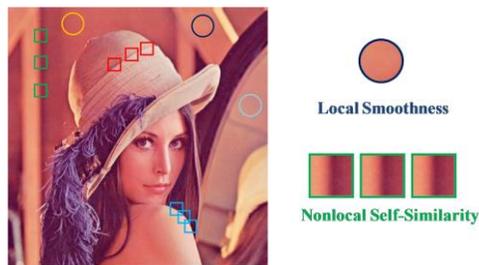

**Fig. 1.** Illustrations for local smoothness and nonlocal self-similarity of natural images.

In this study, we characterize these two properties from the perspective of image statistics and propose a joint statistical modeling (JSM) for high fidelity of image restoration in an adaptive hybrid space-transform domain. Specifically, JSM is established by merging two complementary models– local statistical modeling (LSM) in two-dimensional space domain and nonlocal statistical modeling (NLSM) in three-dimensional transform domain. That is

$$\Psi_{JSM}(\boldsymbol{u}) = \tau \cdot \Psi_{LSM}(\boldsymbol{u}) + \lambda \cdot \Psi_{NLSM}(\boldsymbol{u}), \quad (3)$$

where $\tau, \lambda$ are regularization parameters, which control the trade-off between two competing statistical terms. $\Psi_{LSM}$ corresponds to the above local smoothness prior and keeps image local consistency, suppressing noise effectively, while $\Psi_{NLSM}$ corresponds to the above nonlocal self-similarity prior and maintains image nonlocal consistency, retaining the sharpness and edges effectually. More details on how to design JSM to characterize the above two properties will be provided below.

### A. Local Statistical Modeling for Smoothness in Space Domain

Local smoothness describes the closeness of neighboring pixels in the two-dimensional space domain of images, which means the intensities of the neighboring pixels are quite similar. To characterize the smoothness of images, there exist many models. Here, we mathematically formulate a local statistical modeling for smoothness in two-dimensional space domain. From the view of statistics, a natural image is preferred when its responses for a set of high-passing filters are as small as possible [23], which intuitively implies that images are locally smooth and their derivatives are close to zero.

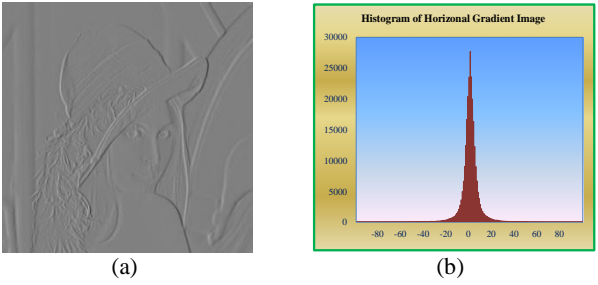

**Fig. 2.** Illustrations for local statistical modeling for smoothness in space domain at pixel level. (a) Gradient picture in horizontal direction of image *Lena*; (b) Distribution of horizontal gradient picture of *Lena*, i.e., histogram of Fig. 2(a).

In practice, the widely-used filters are vertical and horizontal finite difference operators, denoted by $\mathcal{D}_v = [1\ -1]^T$ and $\mathcal{D}_h = [1\ -1]$, respectively. Fig. 2 shows the gradient picture in horizontal direction of image *Lena* and its histogram. It is obvious to see that the distribution is very sharp and most pixels values are near zero. In literatures, the marginal statistics of outputs of the above two filters are usually modeled by generalized Gaussian distribution (GGD) [43], which is defined as

$$p_{\text{GGD}}(x) = \frac{v \cdot \eta(v)}{2 \cdot \Gamma(1/v)} \cdot \frac{1}{\sigma_x} e^{-[\eta(v)|x|/\sigma_x]^v}, \quad (4)$$

where $\eta(v) = \sqrt{\Gamma(3/v)\Gamma(1/v)}$ and $\Gamma(t) = \int_0^\infty e^{-u} u^{t-1} du$ is gamma function, $\sigma_x$ is the standard deviation, and $v$ is the shape parameter. The distribution $p_{\text{GGD}}(x)$ is a Gaussian distribution function if $v=2$, and a Laplacian distribution function if $v=1$. If $0 < v < 1$, $p_{\text{GGD}}(x)$ is named as hyper-Laplacian distribution. More discussions about the value of $v$ can be found in [23].

In this section, we choose Laplacian distribution to model the marginal distributions of gradients of natural images by making a trade-off between modeling the image statistics accurately and being able to solve the ensuing optimization problem efficiently. Thus, let $\mathcal{D} = [\mathcal{D}_v; \mathcal{D}_h]$ and set $v$ to be 1 in Eq. (4) to obtain the local statistical modeling (LSM) in space domain at pixel level, with corresponding regularization term $\Psi_{LSM}$ denoted by

$$\Psi_{LSM}(u) = \|\mathcal{D}u\|_1 = \|\mathcal{D}_v u\|_1 + \|\mathcal{D}_h u\|_1, \quad (5)$$

which clearly indicates that the formulation is convex and facilitates the theoretical analysis.

Note that $\Psi_{LSM}$ has the same expression as anisotropic TV defined in [14], [44], and can be regarded as a statistical interpretation of anisotropic TV. It is important to emphasize that local statistical modeling is only used for characterizing the property of image smoothness. The regularization term Eq. (5) has the advantages of convex optimization and low computational complexity. There is no need to design a very complex regularization term, since the task of retaining the sharp edges and recovering the fine textures will be accomplished by the following nonlocal statistical modeling. More details for solving LSM regularized problems will be given in the next section.

### B. Nonlocal Statistical Modeling for Self-Similarity in Transform Domain

Besides local smoothness, nonlocal self-similarity is another significant property of natural images. It characterizes the repetitiveness of the textures or structures embodied by natural images within nonlocal area, which can be used for retaining the sharpness and edges effectually to maintain image nonlocal consistency. However, the traditional nonlocal regularization terms as mentioned in Section I essentially adopt a weighted manner to characterize self-similarity by introducing nonlocal graph according to the degree of similarity among similar blocks, which often fail to recover finer image textures and more accurate structures.

Recently, quite impressive results have been achieved in image and video denoising by conducting the operation of transforming a three-dimensional (3D) array of similar patches and shrinking the coefficients [4], [32]–[34]. It is worth emphasizing that Dabov *et al.* did excellent work in the image restoration field, especially their famous BM3D methods for image denoising and deblurring applications [4], [21], which have achieved great success. Our proposed statistical modeling for self-similarity is inspired by their success and significantly depends on their work. In this study, we mathematically characterize the nonlocal self-similarity for natural images by means of the distributions of the transform coefficients, which are achieved by transforming the 3D array generated by stacking similar image patches. Accordingly, this type of model can be named as nonlocal statistical modeling (NLSM) for self-similarity in three-dimensional transform domain.

More specifically, as illustrated in Fig. 3, the strict description on the proposed NLSM for self-similarity in transform domain can be obtained in the following five steps. First, divide the image $u$ with size $N$ into $n$ overlapped blocks $u^i$ of size $b_s$, $i=1,2,...,n$. Second, for each block in red denoted by $u^i$, we search $c$ blocks (such as nine in Fig. 3) that are best similar to it within the blue search window. Instead of using a tunable threshold to choose similar blocks in [4] for denoising, our choice with a fixed number is not only simple but also robust to the similarity criterion. Thus, for simplicity, the criterion for calculating similarity between different blocks is Euclidean distance. Moreover, it enables solving the sub-problem associated with NLSM quite efficient (see **Theorem 2**). Define $S_{u^i}$ the set including the $c$ best matched blocks to $u^i$ in the searching window with size of $L \times L$, that is, $S_{u^i} = \{S_{u^i \otimes 1}, S_{u^i \otimes 2},...,S_{u^i \otimes c}\}$. Third, as to each $S_{u^i}$, stack the $c$ blocks belonging to $S_{u^i}$ into a 3D array, which is denoted by $Z_{u^i}$. Fourth, denote $T^{3D}$ the operator of an orthogonal 3D transform, and $T^{3D}(Z_{u^i})$ the transform coefficients for $Z_{u^i}$. Let $\Theta_u$ be the column vector of all the transform coefficients of image $u$ with size $K = b_s * c * n$ built from all the $T^{3D}(Z_{u^i})$ arranged in the lexicographic order. Note that the orthogonality of 3D transform is momentous in solving NLSM, which will be discussed in the next section. Finally, we analyze the histogram of the transform coefficients, as shown in Fig. 3, which statistically demonstrates that the histogram is quite sharp, and the vast majority of coefficients are concentrated near the zero value. This is similar to the previous local modeling of images, and is also very suitable to be characterized by GGD. Analogous to local statistical modeling (LSM) in space domain, by making a trade-off between accurate modeling and efficient solving, in this paper, the distribution of $\Theta_u$ is modeled by Laplacian function.

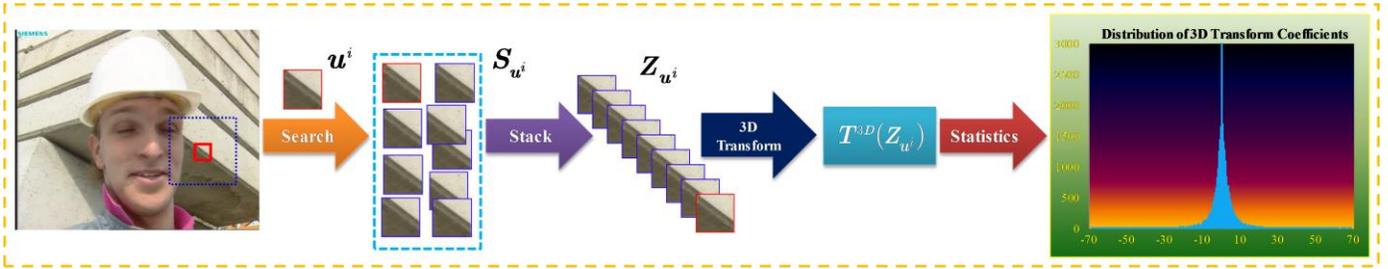

**Fig. 3.** Illustrations for nonlocal statistical modeling for self-similarity in three-dimensional transform domain at block level.

Therefore, the mathematical formulation of nonlocal statistical modeling for self-similarity in three-dimensional transform domain is written as

$$\Psi_{NLSM}(\boldsymbol{u}) = \|\Theta_u\|_1 = \sum_{i=1}^{n} \|\boldsymbol{T}^{3D}(\boldsymbol{Z}_{u^i})\|_1. \quad (6)$$

Accordingly, the inverse operator $\Omega_{NLSM}$ corresponding to $\Psi_{NLSM}$ can be defined in the following procedures. After obtaining $\Theta_u$, split it into $n$ 3D arrays of 3D transform coefficients, which are then inverted to generate estimates for each block in the 3D array. The block-wise estimates are returned to their original positions and the final image estimate is achieved by averaging all of the above block-wise estimates. Forasmuch, if $\Theta_u$ is known, the new estimate for $\boldsymbol{u}$ is expressed as $\hat{\boldsymbol{u}} = \Omega_{NLSM}(\Theta_u)$. The convexity of NLSM in Eq. (6) can be technically justified as follows. To make it clear, define $\boldsymbol{R}_i^{3D}$ as the matrix operator that extracts the 3D array $\boldsymbol{Z}_{u^i}$ from $\boldsymbol{u}$, i.e., $\boldsymbol{Z}_{u^i} = \boldsymbol{R}_i^{3D} \boldsymbol{u}$. Then, define $\boldsymbol{G}_i^{3D} = \boldsymbol{T}^{3D} \boldsymbol{R}_i^{3D}$, which is a linear operator. It is obvious to observe that $\|\boldsymbol{T}^{3D}(\boldsymbol{Z}_{u^i})\|_1 = \|\boldsymbol{G}_i^{3D} \boldsymbol{u}\|_1$ is convex with respect to $\boldsymbol{u}$. Since the sum of convex functions is convex, Eq. (6) is also convex as to $\boldsymbol{u}$.

The difference between the proposed NLSM and BM3D method mainly has three aspects. First, we mathematically characterize the nonlocal self-similarity for natural images by means of the distributions of the transform coefficients, which are achieved by transforming the 3D array generated by stacking similar image blocks. Second, for each block, we utilize a fixed number of blocks that are best similar to it within the search window to construct its 3D array. Nonetheless, in the BM3D works [4], [21], many tunable thresholds to choose similar blocks are exploited, which is a bit complicated. Our choice with a fixed number is not only simple but also robust to the similarity criterion. Moreover, the fixed size of each 3D array enables solving the sub-problem associated with NLSM quite efficient (see **Theorem 2**). Third, the proposed NLSM is more general, and can be directly incorporated into the regularization framework for image inverse problems, such as image inpainting, image deblurring and mixed Gaussian plus impulse noise removal, which will be provided in the experimental section. Furthermore, a split Bregman based iterative algorithm and a theorem are developed to solving the NLSM regularized problem effectively and efficiently.

Here, we also give a visual comparison between the proposed nonlocal statistical modeling (NLSM) and two traditional nonlocal regularization terms. Fig. 4 provides visual results of image restoration from partial random samples for crops of image *Barbara* in the case of *Ratio=20%*. Fig. 4(a) is the corresponding degraded image with only *20%* random samples available. Fig. 4(b) is the reconstruction result achieved only by local statistical modeling (LSM). It looks good in smooth regions, but loses sharp edges and accurate textures. Fig. 4(c) is the reconstruction result achieved by local statistical modeling and NLM-based regularization term together, denoted by LSM+NLM, where the nonlocal regularization term is defined in [45]. Fig. 4(d) provides the restoration result by nonlocal total variation (NL/TV), defined in [28]. It is obvious that reconstruction result with sharper edges and more image details is obtained by incorporation nonlocal graph However, accurate image textures still can't be recovered and the results are not very clear (see the scarf in Fig. 4(d)). Fig. 4(e) shows the restoration result by the proposed local statistical modeling (LSM) plus nonlocal statistical modeling (NLSM), i.e., the proposed JSM.

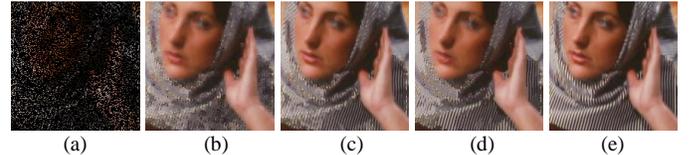

(a)      (b)      (c)      (d)      (e)

**Fig. 4.** Visual quality comparison of image restoration from partial random samples for crops of image *Barbara* in the case of *Ratio=20%*. (a) Degraded image with only *20%* random samples available; (b) Restoration results by only local statistical modeling, i.e., LSM (22.18 dB); (c) Restoration results by LSM+NLM (22.97 dB); (d) Restoration results by NL/TV (23.08 dB); (e) Restoration results by LSM+NLSM, i.e., the proposed JSM (27.21 dB).

It can be observed that Fig. 4(e) exhibits the best visual quality, not only providing consistent and sharp edges but also generating accurate and clear textures, which fully substantiates the superiority of the proposed NLSM over the traditional nonlocal regularizers.

In summary, the advantage of the nonlocal statistical modeling is that self-similarity among globally positioned image blocks is exploited in a more effective statistical manner in 3D transform domain than nonlocal graph incorporated in traditional nonlocal regularizations. Extensive experiments in the following section demonstrate that the NLSM for self-similarity is able to not only reserve the common textures and details among all similar patches, but also keep the distinguished features of each block in a certain degree. Note that the nonlocal statistical modeling for self-similarity is da-ta-adaptive because of its content-aware search for similar blocks within nonlocal region. It is also worth stressing that although Eq. (6) seems complicated as one regularization term in the minimization function, we will give a very efficient solution in the next section.

*C. Joint Statistical Modeling (JSM)*

Considering local smoothness and nonlocal self-similarity in a whole, a new joint statistical modeling (JSM) can be defined by combining the local statistical modeling (LSM) for smoothness in space domain at pixel level and the nonlocal statistical modeling (NLSM) in transform domain at block level, which is expressed as

$$\Psi_{JSM}(\boldsymbol{u}) = \tau \cdot \Psi_{LSM}(\boldsymbol{u}) + \lambda \cdot \Psi_{NLSM}(\boldsymbol{u}) = \tau \cdot \|\mathcal{D}\boldsymbol{u}\|_1 + \lambda \cdot \|\Theta_{\boldsymbol{u}}\|_1. \quad (7)$$

Thus, JSM is able to portray local smoothness and nonlocal self-similarity of natural images richly, and combine the best of the both worlds, which greatly confines the space of inverse problem solution and significantly improve the reconstruction quality. To make JSM tractable and robust, a new Split Bregman based iterative algorithm is developed to solve the optimization problem with JSM as regularization term efficiently, whose implementation details and convergence proof will be provided in the next section. Extensive experimental results will testify the validity of the proposed JSM.

Fig. 5 visually illustrates the image restoration process of the proposed algorithm. Fig. 5(a) is the degraded image of *House* with *20%* original samples, i.e., *Ratio=20%*. As the iteration number *k* increases, it is obvious that the quality of the restoration image becomes better and better, and ultimately stabilizes, exhibited by Figs. 5(b)-(e).

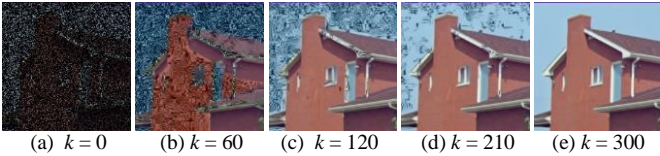

(a) *k* = 0   (b) *k* = 60   (c) *k* = 120   (d) *k* = 210   (e) *k* = 300
**Fig. 5.** Image restoration process as the iteration number increases in the case of image restoration from partial random samples for image *House* when *Ratio=20%*. Here, *k* represents the iteration number.

### III. SPLIT BREGMAN BASED ITERATIVE ALGORITHM FOR IMAGE RESTORATION USING JSM

By incorporating the proposed joint statistical modeling (7) into the regularization-based framework (2), a new formulation for image restoration can be expressed as follows:

$$\operatorname{argmin}_{\boldsymbol{u}} \tfrac{1}{2}\|\boldsymbol{H}\boldsymbol{u} - \boldsymbol{y}\|_2^2 + \tau \cdot \Psi_{LSM}(\boldsymbol{u}) + \lambda \cdot \Psi_{NLSM}(\boldsymbol{u}), \quad (8)$$

where $\tau$ and $\lambda$ are control parameters. Note that the first term of Eq. (8) actually represents the observation constraint and the second and the third represent the image prior local and nonlocal constraints, respectively. Therefore, it is our belief that better results will be achieved by imposing the above three constraints into the ill-posed image inverse problem. Solving it efficiently is one of the main contributions of this paper.

In this section, we apply the algorithmic framework of Split Bregman Iteration to solve Eq. (8) and present the implementation details and the convergence of the proposed algorithm.

Split Bregman Iteration (SBI) is recently introduced by [41] for solving a class of $\ell_1$ related minimization problems. The basic idea of SBI is to convert the unconstrained minimization problem into a constrained one by introducing the variable splitting technique and then invoke the Bregman iteration [41] to solve the constrained minimization problem. Numerical simulations in [40], [44] show that it converges fast and only uses a small memory footprint, which makes it very attractive for large-scale problems.

Consider an unconstrained optimization problem

$$\min_{\boldsymbol{u} \in \mathbb{R}^N} f(\boldsymbol{u}) + g(\boldsymbol{G}\boldsymbol{u}), \quad (9)$$

where $\boldsymbol{G} \in \mathbb{R}^{M \times N}$, $f: \mathbb{R}^N \to \mathbb{R}$, $g: \mathbb{R}^M \to \mathbb{R}$. The Split Bregman Iteration works as follows:

**Algorithm 1** *Split Bregman Iteration (SBI)*
1.     **Set** $k = 0$, choose $\mu > 0$, $\boldsymbol{d}^{(0)} = \boldsymbol{0}, \boldsymbol{u}^{(0)} = \boldsymbol{0}, \boldsymbol{v}^{(0)} = \boldsymbol{0}$.
2.     **Repeat**
3.         $\boldsymbol{u}^{(k+1)} = \operatorname{argmin}_{\boldsymbol{u}} f(\boldsymbol{u}) + \tfrac{\mu}{2}\|\boldsymbol{G}\boldsymbol{u} - \boldsymbol{v}^{(k)} - \boldsymbol{d}^{(k)}\|_2^2$;
4.         $\boldsymbol{v}^{(k+1)} = \operatorname{argmin}_{\boldsymbol{v}} g(\boldsymbol{v}) + \tfrac{\mu}{2}\|\boldsymbol{G}\boldsymbol{u}^{(k+1)} - \boldsymbol{v} - \boldsymbol{d}^{(k)}\|_2^2$;
5.         $\boldsymbol{d}^{(k+1)} = \boldsymbol{d}^{(k)} - (\boldsymbol{G}\boldsymbol{u}^{(k+1)} - \boldsymbol{v}^{(k+1)})$;
6.         $k \leftarrow k + 1$;
7.     **Until** stopping criterion is satisfied

Let us go back to Eq. (8) and point out how to apply SBI to solve it. First, define

$$f(\boldsymbol{u}) = \tfrac{1}{2}\|\boldsymbol{H}\boldsymbol{u} - \boldsymbol{y}\|_2^2,$$
$$g(\boldsymbol{v}) = g(\boldsymbol{G}\boldsymbol{u}) = \tau \cdot \Psi_{LSM}(\boldsymbol{u}) + \lambda \cdot \Psi_{NLSM}(\boldsymbol{u}),$$

where

$$\boldsymbol{v} = \begin{bmatrix} \boldsymbol{w} \\ \boldsymbol{x} \end{bmatrix} = \boldsymbol{G}\boldsymbol{u}, \ \boldsymbol{w}, \boldsymbol{x} \in \mathbb{R}^N \text{ and } \boldsymbol{G} = \begin{bmatrix} \boldsymbol{I} \\ \boldsymbol{I} \end{bmatrix} \in \mathbb{R}^{2N \times N}.$$

Therefore, Eq. (8) is transformed to

$$\operatorname{argmin}_{\boldsymbol{u} \in \mathbb{R}^N, \boldsymbol{v} \in \mathbb{R}^{2N}} f(\boldsymbol{u}) + g(\boldsymbol{v}) \text{ s. t. } \boldsymbol{G}\boldsymbol{u} = \boldsymbol{v}. \quad (10)$$

Invoking SBI, Line 3 in Algorithm 1 becomes:

$$\begin{aligned}
\boldsymbol{u}^{(k+1)} &= \operatorname*{argmin}_{\boldsymbol{u}} f(\boldsymbol{u}) + \tfrac{\mu}{2}\|\boldsymbol{G}\boldsymbol{u} - \boldsymbol{v}^{(k)} - \boldsymbol{d}^{(k)}\|_2^2 \\
&= \tfrac{1}{2}\|\boldsymbol{H}\boldsymbol{u} - \boldsymbol{y}\|_2^2 + \tfrac{\mu}{2}\left\|\begin{bmatrix}\boldsymbol{I}\\\boldsymbol{I}\end{bmatrix}\boldsymbol{u} - \begin{bmatrix}\boldsymbol{w}^{(k)}\\\boldsymbol{x}^{(k)}\end{bmatrix} - \begin{bmatrix}\boldsymbol{b}^{(k)}\\\boldsymbol{c}^{(k)}\end{bmatrix}\right\|_2^2,
\end{aligned} \quad (11)$$

where $\boldsymbol{d}^{(k)} = \begin{bmatrix} \boldsymbol{b}^{(k)} \\ \boldsymbol{c}^{(k)} \end{bmatrix} \in \mathbb{R}^{2N}$, $\boldsymbol{b}^{(k)}, \boldsymbol{c}^{(k)} \in \mathbb{R}^N$.

Splitting $\ell_2$ norm in Eq. (11), we have

$$\begin{aligned}
\boldsymbol{u}^{(k+1)} = \operatorname*{argmin}_{\boldsymbol{u}} &\tfrac{1}{2}\|\boldsymbol{H}\boldsymbol{u} - \boldsymbol{y}\|_2^2 + \tfrac{\mu}{2}\|\boldsymbol{u} - \boldsymbol{w}^{(k)} - \boldsymbol{b}^{(k)}\|_2^2 \\
&+ \tfrac{\mu}{2}\|\boldsymbol{u} - \boldsymbol{x}^{(k)} - \boldsymbol{c}^{(k)}\|_2^2.
\end{aligned} \quad (12)$$

Next, Line 4 in **Algorithm 1** becomes:

$$\begin{aligned}
\boldsymbol{v}^{(k+1)} = \begin{bmatrix}\boldsymbol{w}^{(k+1)}\\\boldsymbol{x}^{(k+1)}\end{bmatrix} = \operatorname*{argmin}_{\boldsymbol{w},\boldsymbol{x}} \big\{ &\tau \cdot \Psi_{LSM}(\boldsymbol{w}) + \lambda \cdot \Psi_{NLSM}(\boldsymbol{x}) \\
&+ \tfrac{\mu}{2}\|\boldsymbol{u}^{(k+1)} - \boldsymbol{w} - \boldsymbol{b}^{(k)}\|_2^2 + \tfrac{\mu}{2}\|\boldsymbol{u}^{(k+1)} - \boldsymbol{x} - \boldsymbol{c}^{(k)}\|_2^2 \big\}.
\end{aligned} \quad (13)$$

Clearly, the minimization with respect to $w$, $x$ are decoupled, thus can be solved separately, leading to

$$w^{(k+1)} = \underset{w}{\operatorname{argmin}} \tau \cdot \Psi_{LSM}(w) + \frac{\mu}{2} \|u^{(k+1)} - w - b^{(k)}\|_2^2, \quad (14)$$

$$x^{(k+1)} = \underset{x}{\operatorname{argmin}} \lambda \cdot \Psi_{NLSM}(x) + \frac{\mu}{2} \|u^{(k+1)} - x - c^{(k)}\|_2^2. \quad (15)$$

According to Line 5 in **Algorithm 1**, the update of $d_k$ is

$$d^{(k+1)} = \begin{bmatrix} b^{(k+1)} \\ c^{(k+1)} \end{bmatrix} = \begin{bmatrix} b^{(k)} \\ c^{(k)} \end{bmatrix} - \left( \begin{bmatrix} I \\ I \end{bmatrix} u^{(k+1)} - \begin{bmatrix} w^{(k+1)} \\ x^{(k+1)} \end{bmatrix} \right),$$

which can be simplified into the following two expresstions:

$$b^{(k+1)} = b^{(k)} - (u^{(k+1)} - w^{(k+1)}),$$
$$c^{(k+1)} = c^{(k)} - (u^{(k+1)} - x^{(k+1)}).$$

To sum up, the minimization for Eq. (8) is equivalent to solve the three sub-problems, namely, $u$, $w$, $x$ sub-problems, according to Split Bregman Iteration. The complete algorithm for solving Eq. (8) is described in Table I.

In Table I, the proximal map $prox_t(g)(x)$ with respect to a proper closed convex function $g$ and a scalar $t > 0$ is defined by $prox_t(g)(x) = \underset{u}{\operatorname{argmin}} \left\{ \frac{1}{2} \|u - x\|_2^2 + t \cdot g(u) \right\}$ [14].

In the light of the convergence of SBI, we have the following theorem to prove the convergence of the proposed algorithm using joint statistical modeling in Table I.

**THEOREM 1.** *The proposed algorithm described by Table I converges to a solution of (8).*
*Proof:* It is obvious that the proposed algorithm is an instance of Split Bregman Iteration. Since all the three functions $f(\cdot)$, $\Psi_{LSM}(\cdot)$, and $\Psi_{NLSM}(\cdot)$ are closed, proper and convex, the convergence of the proposed algorithm is guaranteed by

$$G = \begin{bmatrix} I \\ I \end{bmatrix} \in \mathbb{R}^{2N \times N},$$

which is a full column rank matrix. □

TABLE I. A COMPLETE DESCRIPTION OF PROPOSED ALGORITHM USING JOINT STATISTICAL MODELING (VERSION I)

**Input:** the observed image $y$ and the linear matrix operator $H$
**Initialization:** $k = 0, u^{(0)} = y, b^{(0)} = c^{(0)} = w^{(0)} = x^{(0)} = 0, \mu, \tau, \lambda$
**Repeat**

$$u^{(k+1)} = \underset{u}{\operatorname{argmin}} \frac{1}{2} \|Hu - y\|_2^2$$
$$+ \frac{\mu}{2} \|u - w^{(k)} - b^{(k)}\|_2^2 + \frac{\mu}{2} \|u - x^{(k)} - c^{(k)}\|_2^2;$$
$$p^{(k)} = u^{(k+1)} - b^{(k)}; \quad \gamma = \tau / \mu;$$
$$w^{(k+1)} = prox_\gamma (\Psi_{LSM})(p^{(k)});$$
$$r^{(k)} = u^{(k+1)} - c^{(k)}; \quad \alpha = \lambda / \mu;$$
$$x^{(k+1)} = prox_\alpha (\Psi_{NLSM})(r^{(k)});$$
$$b^{(k+1)} = b^{(k)} - (u^{(k+1)} - w^{(k+1)});$$
$$c^{(k+1)} = c^{(k)} - (u^{(k+1)} - x^{(k+1)});$$

**Until** stopping criterion is satisfied
**Output:** Final restored image $u$.

It is important to stress that the convergence will not be compromised if the sub-problems can be solved efficiently, which will also be demonstrated by the following experimental section. In the following, we argue that the every separated sub-problem admits an efficient solution. For simplicity, the subscript $k$ is omitted without confusion.

### A. $u$ sub-problem

In order to make the solution of Eq. (12) more flexible, we introduce two parameters $\mu_1$ and $\mu_2$ to replace $\mu$, which will not comprise the algorithm convergence. Thus, given $w, x$, the $u$ sub-problem denoted by Eq. (12) becomes:

$$u = \underset{u}{\operatorname{argmin}} \frac{1}{2} \|Hu - y\|_2^2 + \frac{\mu_1}{2} \|u - w - b\|_2^2 + \frac{\mu_2}{2} \|u - x - c\|_2^2. \quad (16)$$

Since Eq. (16) is a minimization problem of strictly convex quadratic function, there is actually a closed form for $u$, which can be expressed as

$$u = (H^T H + \tilde{\mu} I)^{-1} \cdot q, \quad (17)$$

where $q = H^T y + \mu_1 (w + b) + \mu_2 (x + c)$, $I$ is identity matrix and $\tilde{\mu} = \mu_1 + \mu_2$. For image inpainting and image deblurring problems, Eq. (17) can be computed efficiently [15].

As for image inpainting, since the sub-sampling matrix $H$ is actually a binary matrix, which can be generated by taking a subset of rows of an identity matrix, $H$ satisfies $HH^T = I$. Applying the Sherman-Morrison-Woodbury (SMW) matrix inversion formula to Eq. (17) yields

$$u = \frac{1}{\tilde{\mu}} (I - \frac{1}{1+\tilde{\mu}} H^T H) \cdot q. \quad (18)$$

Therefore, $u$ in Eq. (18) can be computed very efficiently without computing the matrix inverse operation in Eq. (17). Moreover, owing to the particular structure of $H$, $H^T H$ is equal to an identity matrix with some zeros in the diagonal, corresponding to the positions of the missing pixels. Consequently, the cost of Eq. (18) is only $O(N)$. In this paper, the mixed Gaussian plus salt-and-pepper noise removal is dealt with as a special case of image inpainting, which will be elaborated in the following section.

As for image deblurring, $H$ is the matrix representing a circular convolution which can be factorized as

$$H = U^{-1} D U, \quad (19)$$

where $U$ is the matrix denoting 2D discrete Fourier transform (DFT), $U^{-1}$ is its inverse and $D$ is a diagonal matrix containing the DFT coefficients of the convolution operator represented by $H$. Thus,

$$(H^T H + \tilde{\mu} I)^{-1} = (U^{-1} D^* D U + \tilde{\mu} U^{-1} U)^{-1} = U^{-1} (|D|^2 + \tilde{\mu} I)^{-1} U, \quad (20)$$

where $(\cdot)^*$ denotes complex conjugate and $|D|^2$ the squared absolute values of the entries of the diagonal matrix $D$. Because $|D|^2 + \tilde{\mu} I$ is diagonal, the cost of its inversion is $O(N)$. In practice, the products of $U^{-1}$ and $U$ can be implemented with $O(N \log N)$ using the FFT algorithm.

### B. $w$ sub-problem

$w$ sub-problem, the proximal map associated to $\Psi_{LSM}(\cdot)$, can be regarded as a denoising filtering with anisotropic total variation as mentioned before. To solve it, one of the intrinsic difficulties is the non-smoothness of the term $\|\mathcal{D}u\|_1$. To overcome this difficulty, Chambolle [3] suggested to consider a dual approach, and developed a globally convergent gradi-ent-based algorithm for the denoising problem, which was shown to be faster than primal-based schemes. Later, some accelerating methods such as TwIST [13] and FISTA [14], are proposed, exhibiting fast theoretical and practical convergence. In our experiments, we exploit a fixed number of iterations of FISTA to solve $w$ sub-problem, which is computationally efficient and empirically found not to compromise convergence of the proposed algorithm.

## C. $x$ sub-problem

Given $w, u$, the $x$ sub-problem can be written as

$$\begin{aligned} x &= prox_\alpha(\Psi_{NLSM})(r) \\ &= \operatorname*{argmin}_x \left\{ \tfrac{1}{2} \|x-r\|_2^2 + \alpha \cdot \Psi_{NLSM}(x) \right\} \\ &= \operatorname*{argmin}_x \left\{ \tfrac{1}{2} \|x-r\|_2^2 + \alpha \|\Theta_x\|_1 \right\}. \end{aligned} \quad (21)$$

By viewing $r$ as some type of noisy observation of $x$, we perform some experiments to investigate the statistics of $e = x - r$. Here, we use color image *Butterfly* as an example in the case of image deblurring, where the original image is first blurred by Gaussian blur kernel and then is added by Gaussian white noise of standard deviation 0.5. At each iteration $t$, we can obtain $r^{(k)}$ by $r^{(k)} = u^{(k)} - c^{(k-1)}$. Since the exact minimizer of Eq. (21) is not available, we then approximate $x^{(k)}$ by the original image without generality. Therefore, we are able to acquire the histogram of $e^{(k)} = x^{(k)} - r^{(k)}$ at each iteration $k$. Fig. 6 shows the distributions of $e^{(k)}$ when $k$ equals to 4 and 8, respectively.

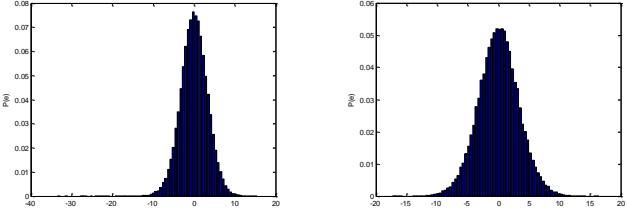

**Fig. 6.** The distribution of $e^{(k)}$ and its corresponding variance $\mathrm{Var}(e^{(k)})$ for image *Butterfly* in the case of image deblurring at different iterations. (a) $k = 4$ and $\mathrm{Var}(e^{(4)}) = 11.18$; (b) $k = 8$ and $\mathrm{Var}(e^{(8)}) = 10.95$.

From Fig. 6, it is obvious to observe that the distribution of $e^{(k)}$ at each iteration is quite suitable to be characterized by generalized Gaussian distribution (GGD) [43] with zero-mean and variance $\mathrm{Var}(e^{(k)})$. The variance $\mathrm{Var}(e^{(k)})$ can be estimated by

$$\mathrm{Var}(e^{(k)}) = \tfrac{1}{N} \|x^{(k)} - r^{(k)}\|_2^2. \quad (22)$$

Fig. 6 also gives the corresponding estimated variances at different iterations. Furthermore, owing that the residual of images is usually de-correlated, each element of $e^{(k)}$ can be modeled independently.

Accordingly, to enable solving Eq. (21) tractable, in this paper, a reasonable assumption is made, with which even a closed-form solution of Eq. (21) can be obtained. We suppose that each element of $e^{(k)}$ follows an independent zero-mean distribution with variance $\mathrm{Var}(e^{(k)})$. It is worth emphasizing that the above assumption does not need to be Gaussian, or Laplacian, or GGD process, which is more general. By this assumption, we can prove the following conclusion.

**THEOREM 2.** Let $x, r \in \mathbb{R}^N, \Theta_x, \Theta_r \in \mathbb{R}^K$, and denote the error vector by $e = x - r$ and each element of $e$ by $e(j)$, $j = 1, ..., N$. Assume that $e(j)$ is independent and comes from a distribution with zero mean and variance $\sigma^2$. Then, for any $\varepsilon > 0$, we have the following property to describe the relationship between $\|x - r\|_2^2$ and $\|\Theta_x - \Theta_r\|_2^2$, that is,

$$\lim_{N \to \infty, K \to \infty} P\left\{ \left| \tfrac{1}{N} \|x-r\|_2^2 - \tfrac{1}{K} \|\Theta_x - \Theta_r\|_2^2 \right| < \varepsilon \right\} = 1, \quad (23)$$

where $P(\cdot)$ represents the probability.

*Proof*: Due to the assumption that each $e(j)$ is independent, we obtain that each $e(j)^2$ is also independent. Since $E[e(j)] = 0$ and $D[e(j)] = \sigma^2$, we have the mean of each $e(j)^2$, which is expressed as

$$E[e(j)^2] = D[e(j)] + [E[e(j)]]^2 = \sigma^2, \quad j = 1, ..., N.$$

By invoking the *Law of Large Numbers* in probability theory, for any $\varepsilon > 0$, it leads to $\lim_{N \to \infty} P\left\{ \left| \tfrac{1}{N} \sum_{j=1}^N e(j)^2 - \sigma^2 \right| < \tfrac{\varepsilon}{2} \right\} = 1$, i.e.,

$$\lim_{N \to \infty} P\left\{ \left| \tfrac{1}{N} \|x-r\|_2^2 - \sigma^2 \right| < \tfrac{\varepsilon}{2} \right\} = 1. \quad (24)$$

Further, denote each element of $\Theta_e$ by $\Theta_e(j), j = 1, ..., K$. Due to the definition of 3D transform coefficients vector $\Theta_e$ and the orthogonal property of transform $T^{3D}$, we conclude that $\Theta_e(j)$ is independent with zero mean and variance $\sigma^2$.

Therefore, the same manipulations applied to $\Theta_e(j)^2$ yield $\lim_{K \to \infty} P\left\{ \left| \tfrac{1}{K} \sum_{j=1}^K \Theta_e(j)^2 - \sigma^2 \right| < \tfrac{\varepsilon}{2} \right\} = 1$, namely,

$$\lim_{K \to \infty} P\left\{ \left| \tfrac{1}{K} \|\Theta_x - \Theta_r\|_2^2 - \sigma^2 \right| < \tfrac{\varepsilon}{2} \right\} = 1. \quad (25)$$

Considering Eqs. (24) and (25) together, we prove Eq. (23). □

According to **Theorem 2**, there exists the following equation with very large probability (limited to 1) at each iteration $k$:

$$\tfrac{1}{N} \|x^{(k)} - r^{(k)}\|_2^2 = \tfrac{1}{K} \|\Theta_x^{(k)} - \Theta_r^{(k)}\|_2^2. \quad (26)$$

Now let's verify Eq. (26) by the above case of image deblurring. We can clearly see that the left hand of Eq. (26) is just $\mathrm{Var}(e^{(k)})$ defined in Eq. (22), with $\mathrm{Var}(e^{(4)}) = 11.18$ and $\mathrm{Var}(e^{(8)}) = 10.95$, which is shown in Fig. 6.

At the same time, we can calculate the corresponding right hand of Eq. (26), denoted by $\mathrm{Var}(\Theta_e^{(k)})$, with the same values of $k$, leading to $\mathrm{Var}(\Theta_e^{(4)}) = 10.98$ and $\mathrm{Var}(\Theta_e^{(8)}) = 10.87$. Apparently, at each iteration, $\mathrm{Var}(e^{(k)})$ is very close to $\mathrm{Var}(\Theta_e^{(k)})$, especially when $k$ is larger, which sufficiently illustrates the validity of our assumption.

Incorporating Eq. (26) into Eq. (21) leads to

$$\operatorname*{argmin}_x \tfrac{1}{2} \|\Theta_x - \Theta_r\|_2^2 + \tfrac{K\alpha}{N} \|\Theta_x\|_1. \quad (27)$$

Since the unknown variable $\Theta_x$ is component-wise separable in Eq. (27), each of its components $\Theta_x(j)$ can be independently obtained in a closed form according to the so called soft thresholding [42]:

$$\Theta_x = \mathbf{soft}(\Theta_r, \sqrt{2\rho}), \quad (28)$$

where $j = 1, ..., K$, $\rho = \tfrac{K\alpha}{N}$ and

$$\Theta_x(j) = \text{sgn}(\Theta_r(j))\max\{|\Theta_r(j)| - \sqrt{2\rho}, 0\}$$

$$= \begin{cases} \Theta_r(j) - \sqrt{2\rho}, & \Theta_r(j) \in (\sqrt{2\rho}, +\infty), \\ 0, & \Theta_r(j) \in [-\sqrt{2\rho}, \sqrt{2\rho}], \\ \Theta_r(j) + \sqrt{2\rho}, & \Theta_r(j) \in (-\infty, -\sqrt{2\rho}). \end{cases}$$

Thus, the closed solution form of $\boldsymbol{x}$ sub-problem Eq. (21) is

$$\boldsymbol{x} = \Omega_{NLSM}(\boldsymbol{\Theta_x}) = \Omega_{NLSM}(\text{soft}(\boldsymbol{\Theta_r}, \sqrt{2\rho})). \quad (29)$$

### D. Summary of Proposed Algorithm

So far, all issues in the process of handing the three sub-problems have been solved efficiently and effectively. In light of all derivations above, a detailed description of the proposed algorithm for image restoration using joint statistical modeling is provided in Table II.

TABLE II. A Complete description of Proposed Algorithm using Joint Statistical Modeling (version II)

**Input:** the observed image $\boldsymbol{y}$ and the linear matrix operator $\boldsymbol{H}$
**Initialization:** $k=0, \boldsymbol{u}^{(0)}=\boldsymbol{y}, \boldsymbol{b}^{(0)}=\boldsymbol{c}^{(0)}=\boldsymbol{w}^{(0)}=\boldsymbol{x}^{(0)}=\boldsymbol{0}, \tau, \lambda, \mu_1, \mu_2$;
**Repeat**
  Compute $\boldsymbol{u}^{(k+1)}$ by Eq. (18) or Eq. (20);
  $\boldsymbol{p}^{(k)} = \boldsymbol{u}^{(k+1)} - \boldsymbol{b}^{(k)}; \gamma = \tau/\mu_1;$
  Compute $\boldsymbol{w}^{(k+1)}$ by FISTA;
  $\boldsymbol{r}^{(k+1)} = \boldsymbol{u}^{(k+1)} - \boldsymbol{c}^{(k+1)}; \alpha = \lambda/\mu_2;$
  Compute $\boldsymbol{x}^{(k+1)}$ by Eq. (29);
  $\boldsymbol{b}^{(k+1)} = \boldsymbol{b}^{(k)} - (\boldsymbol{u}^{(k+1)} - \boldsymbol{w}^{(k+1)});$
  $\boldsymbol{c}^{(k+1)} = \boldsymbol{c}^{(k)} - (\boldsymbol{u}^{(k+1)} - \boldsymbol{x}^{(k+1)});$
**Until** maximum iteration number is reached
**Output:** Final restored image $\boldsymbol{u}$.

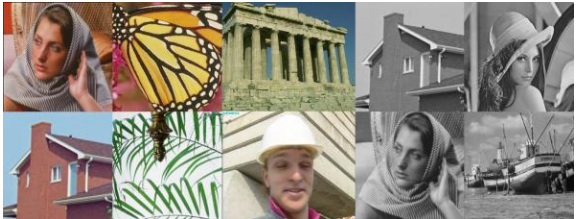
**Fig. 7.** All experimental test images.

## IV. Experimental Results

In this section, extensive experimental results are presented to evaluate the performance of the proposed algorithm, which is compared with many state-of-the-art methods. We apply our algorithm to the applications of image inpainting, image deblurring and mixed Gaussian plus salt-and-pepper noise removal. All the experiments are performed in Matlab 7.12.0 on a Dell OPTIPLEX computer with Intel(R) Core(TM) 2 Duo CPU E8400 processor (3.00GHz), 3.25G memory, and Windows XP operating system. In our implementation, if not specially stated, the size of each block, i.e., $b_s$ is set to be $8 \times 8$ with 4-pixel-width between adjacent blocks, the size of training window for searching matched blocks, i.e., $L \times L$ is set to be $40 \times 40$, and the number of best matched blocks, i.e., $c$ is set to be 10. Thus, the relationship between $N$ and $K$ is $K = 40N$. The orthogonal 3D transform denoted by $\boldsymbol{T}^{3D}$ is composed of 2D discrete cosine transform and 1D Haar transform. All experimental images are shown in Fig. 7.

To evaluate the quality of image reconstruction, in addition to PSNR (Peak Signal to Noise Ratio, unit: dB), which is used to evaluate the objective image quality, a new image quality assessment (IQA) model FSIM is exploited to evaluate the visual quality. FSIM is proposed recently and achieves much higher consistency with the subjective evaluations than state-of-the-art IQA metrics [31]. The higher FSIM value means the better visual quality, while the FSIM value lies in the interval [0 1]. Note that the results of every color image are obtained by its luminance component, keeping its chrominance components unchanged. In the following, the left of the slash denotes PSNR (dB) and the right of the slash denotes FSIM. Due to the limit of space, only parts of the experimental results are shown in this paper. **Please enlarge and view the figures on the screen for better comparison.** Our Matlab software and more experimental visual results can be downloaded at the website: http://idm.pku.edu.cn/staff/zhangjian/IRJSM/.

### A. Image Restoration from Partial Random Samples

We now handle the problem of image restoration from partial random samples, where the original image is operated by a random mask and the random mask is assumed to be known. That means $\boldsymbol{H}$ in Eq. (8) is already known. The proposed algorithm is compared with five recent representative methods: SKR (steering kernel regression) [35], FoE (fields of experts) [36], MCA (morphological component analysis) [37] and SALSA [15] and BPFA [38].

SKR utilizes a steering kernel regression framework to characterize local structures for image restoration [35]. MCA calculates the sparse inverse problem estimate in a dictionary that combines a curvelet frame, a wavelet frame and a local DCT basis [37]. FoE learns a Markov random field model, where the parameters are trained from huge amounts of example natural images [36]. SALSA develops a fast algorithm for total variation regularization [15]. BPFA exploits the beta process factor analysis framework to infer a learned dictionary using the truncated beta-Bernoulli process [38]. The results of the five comparative methods are generated by the original authors' softwares, with the parameters manually optimized.

Here, three color images are tested, with the percentage of retaining original samples, denoted by *Ratio*, being *20%, 30%, 50%* and *80%*, respectively. The maximum iteration number in Table II is dependent on *Ratio*. In our experiment, the maximum iteration number is set to be 400, 350, 250, and 100 for the above four *Ratios*.

Table III lists PSNR/FSIM results among different methods on the test images. From Table III, the proposed method achieves the highest scores of PSNR and FSIM in all the cases, which fully demonstrates that the restoration results by the proposed method are the best both objectively and visually.

More specifically, the proposed algorithm obtains PSNR improvement of about 2.7 dB and FSIM improvement of about 0.016 on average over the second best algorithms (i.e., BPFA). Note that, in the case of *Ratio=20%* on i*mage House*, the average PSNR and FSIM improvements achieved by the proposed method over BPFA is 4.2 dB and 0.02, separately.

TABLE III. PSNR/FSIM Comparisons of Various Methods for Image Restoration from Partial Random Samples

| Image | Ratio | Degraded | SALSA [15] | MCA [37] | SKR [35] | BPFA [38] | FoE [36] | Proposed |
|---|---|---|---|---|---|---|---|---|
| *House* | 20% | 6.16/0.3962 | 29.17/0.8916 | 32.22/0.9320 | 30.40/0.9198 | 30.89/0.9111 | 32.65/0.9335 | **35.11/0.9564** |
| | 30% | 6.74/0.3840 | 31.53/0.9300 | 34.92/0.9563 | 32.48/0.9518 | 33.84/0.9484 | 35.06/0.9593 | **37.85/0.9741** |
| | 50% | 8.22/0.3854 | 35.00/0.9685 | 38.54/0.9786 | 36.86/0.9784 | 39.57/0.9817 | 39.04/0.9831 | **41.24/0.9879** |
| | 80% | 12.19/0.5210 | 41.03/0.9925 | 43.89/0.9936 | 44.75/0.9954 | 44.10/0.9931 | 45.62/0.9961 | **47.39/0.9971** |
| *Barbara* | 20% | 7.36/0.4998 | 22.75/0.8193 | 25.69/0.8939 | 21.92/0.8607 | 25.70/0.8927 | 23.68/0.8812 | **27.54/0.9264** |
| | 30% | 7.94/0.5111 | 23.65/0.8684 | 27.97/0.9279 | 23.42/0.9042 | 28.44/0.9362 | 25.66/0.9192 | **31.06/0.9618** |
| | 50% | 9.43/0.5292 | 25.93/0.9303 | 32.35/0.9686 | 29.12/0.9639 | 34.18/0.9772 | 30.52/0.9676 | **36.77/0.9864** |
| | 80% | 13.36/0.6484 | 31.82/0.9827 | 40.20/0.9937 | 39.95/0.9947 | 41.25/0.9942 | 39.84/0.9944 | **44.30/0.9973** |
| *Foreman* | 20% | 4.57/0.3551 | 26.27/0.9065 | 31.40/0.9480 | 30.35/0.9492 | 29.64/0.9298 | 30.80/0.9397 | **33.28/0.9631** |
| | 30% | 5.14/0.3295 | 28.41/0.9353 | 33.21/0.9630 | 31.75/0.9683 | 32.01/0.9560 | 33.00/0.9598 | **35.33/0.9745** |
| | 50% | 6.61/0.3209 | 32.46/0.9702 | 36.10/0.9797 | 35.91/0.9852 | 36.73/0.9818 | 36.72/0.9820 | **38.65/0.9880** |
| | 80% | 10.54/0.4498 | 38.91/0.9926 | 41.70/0.9936 | 43.15/0.9961 | 44.13/0.9957 | 43.82/0.9961 | **44.43/0.9970** |
| Avg. | | 8.19/0.4442 | 30.58/0.9323 | 34.85/0.96072 | 33.34/0.9556 | 35.05/0.9582 | 34.70/0.9593 | **37.75/0.9758** |

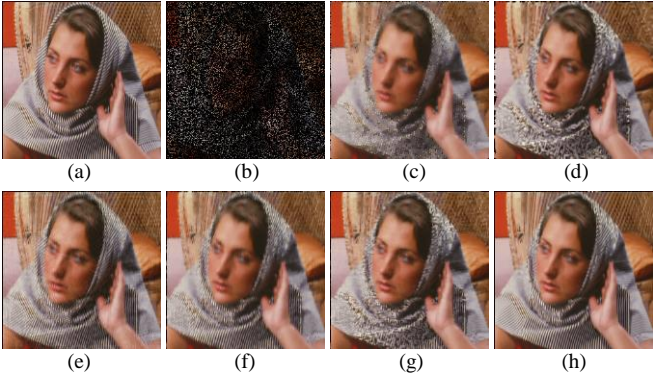

(a) (b) (c) (d)
(e) (f) (g) (h)

**Fig. 8.** Visual quality comparison of image restoration from partial random samples for image *Barbara* in the case of *Ratio=20%*. (a) Original image; (b) Degraded image with only *20%* random samples available (7.36 dB/0.4998); (c)–(h) Restoration results by SALSA (22.75 dB/0.8193) [15], SKR (21.92 dB/0.8607) [35], MCA (25.69 dB/0.8939) [37], BPFA (25.70 dB/0.8927) [38], FoE (23.68 dB/0.8812) [36], and the proposed algorithm (27.54 dB/0.9264).

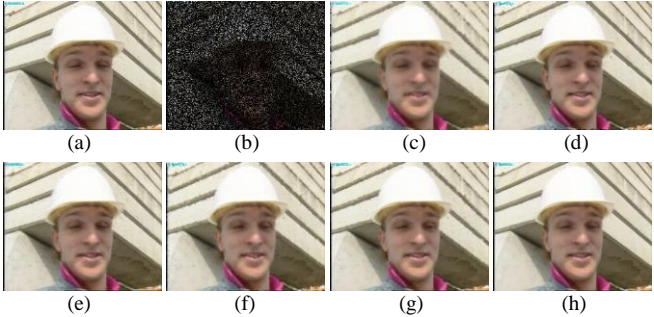

(a) (b) (c) (d)
(e) (f) (g) (h)

**Fig. 9.** Visual quality comparison of image restoration from partial random samples for image *Foreman* in the case of *Ratio=20%*. (a) Original image; (b) Degraded image with only *20%* random samples available (4.57 dB/0.3551); (c)–(h) Restoration results by SALSA (26.27 dB/0.9065) [15], SKR (30.35 dB/0.9492) [35], MCA (31.40 dB/0.9480) [37], BPFA (29.64 dB/0.9298) [38], FoE (30.80 dB/0.9397) [36], and the proposed algorithm (33.28 dB/0.9631).

Figs. 8–9 show visual quality restoration results for *Barbara* and *Foreman* in the case of *Ratio=20%*, where the degraded images (i.e., Figs. 8(b)–9(b)) are hardly identified. It is apparent that all the methods generate good results on the smooth regions. SKR [35] is good at capturing contour structures, but fails in recovering textures and produces blurred effects. MCA [37] can restore better textures than FoE [36] and SKR. However, it produces noticeable striped artifacts. BPFA [38] is able to recover some textures, while generating some incorrect textures and some blurred effects due to less robustness with so small percentage of retaining samples for dictionary learning. The proposed joint statistical modeling (JSM) not only provides accurate restoration on both edges and textures but also suppresses the noise-caused artifacts, exhibiting the best visual quality, which is consistent with FSIM.

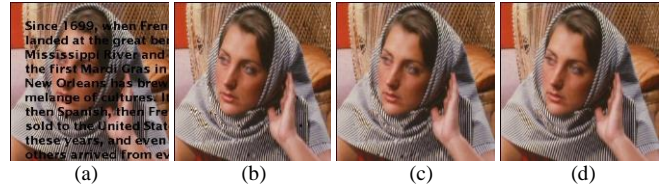

(a) (b) (c) (d)

**Fig. 10.** Visual quality comparison of text removal for image *Barbara*. (a) Degraded image with text mask (15.03 dB/0.7266); (b)–(d) Restoration results by SKR (30.93 dB/0.747) [35], FoE (31.53 dB/0.9745) [36], and the proposed algorithm (37.99 dB/0.9899).

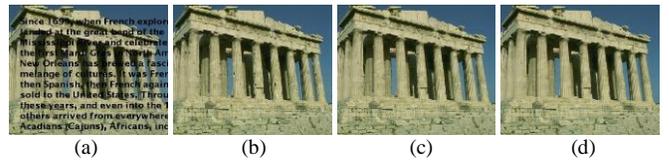

(a) (b) (c) (d)

**Fig. 11.** Visual quality comparison of text removal for image *Parthenon*. (a) Degraded image with text mask (13.91 dB/0.7213); (b)–(d) Restoration results by SKR (31.02 dB/0.9666) [35], FoE (33.23 dB/0.9704) [36], and the proposed algorithm (34.45 dB/0.9770).

### B. Image Restoration for Text Removal

We now deal with another interesting case of image inpainting, i.e., text removal. That means $H$ is not a random mask, but a text one. Four color images are degraded by a known text mask. The purpose for text removal is to infer original images from the degraded versions by removing the text region. The proposed algorithm is compared with three state-of-the-art approaches: SKR [35], FoE [36], and BPFA [38]. The experimental setting for text removal of our proposed algorithm is the same as the one for image restoration from partial random samples. Table IV lists the PSNR and FSIM results among different methods on test images. It shows that the proposed algorithm achieves the highest values in all the cases, which substantiates the effectiveness of the proposed algorithm. Figs. 10–11 further visually illustrates that the proposed algorithm provides more accurate edges and textures with better visual quality, compared with other methods.

TABLE IV. PSNR/FSIM Comparisons for Text Removal

| Image | Barbara | Foreman | House | Parthenon |
|---|---|---|---|---|
| Degraded | 15.03/0.7266 | 11.98/0.6439 | 14.20/0.6499 | 13.91/0.7213 |
| SKR [35] | 30.93/0.9747 | 40.40/0.9930 | 38.65/0.9850 | 31.02/0.9666 |
| FoE [36] | 31.53/0.9745 | 40.39/0.9911 | 39.46/0.9845 | 33.23/0.9704 |
| BPFA [38] | 34.27/0.9790 | 41.09/0.9906 | 38.97/0.9818 | 33.26/0.9697 |
| Proposed | 37.99/0.9899 | 44.92/0.9946 | 41.91/0.9905 | 34.45/0.9770 |

### C. Image Deblurring

In the case of image deblurring, the original images are blurred by a blur kernel and then added by Gaussian noise with standard deviation $\sigma$. Three blur kernels, a 9×9 uniform kernel, a Gaussian blur kernel and a motion blur kernel, are exploited for simulation (see Table VI). We compare the proposed JSM deblurring method to three recently developed deblurring approaches, i.e., the constrained TV deblurring (denoted by SALSA) method [15], the SA-DCT deblurring method [12], and the BM3D deblurring method [21]. Note that SALSA is a recently proposed TV-based deblurring method that can reconstruct the piecewise smooth regions. The SA-DCT and BM3D are two well-known image restoration methods that often produce state-of-the-art image deblurring results.

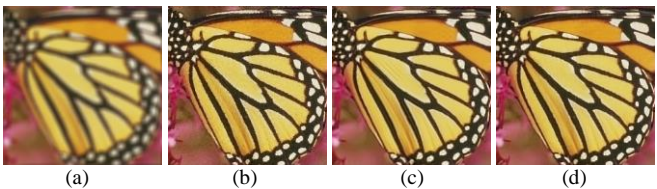

**Fig. 12.** Visual quality comparison of image deblurring on image *Butterfly* (9×9 uniform blur). (a) Noisy and blurred; (b) SALSA (30.30 dB/0.9300) [15]; (c) BM3D (28.73 dB/0.8959) [21]; (d) Proposed (31.03 dB/0.9394).

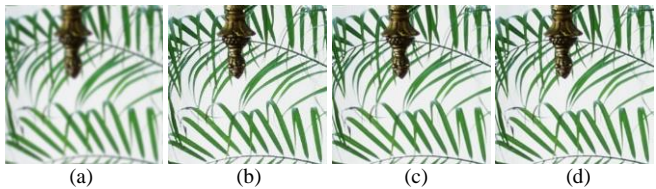

**Fig. 13.** Visual quality comparison of image deblurring on image *Leaves* (Gaussian blur). (a) Noisy and blurred; (b) SALSA (30.32 dB/0.9518) [15]; (c) BM3D (30.61 dB/0.9342) [21]; (d) Proposed (32.18 dB/0.9610).

The PSNR and FSIM results on a set of four images are reported in Table V. From Table V, we can conclude that the proposed JSM approach significantly outperforms other competing methods for all three types of blur kernels. The visual comparisons of the deblurring methods are shown in Fig. 12 and Fig. 13, from which one can observe that the JSM model produces much cleaner and sharper image edges and textures than other methods with almost unnoticeable ringing artifacts. The high performance of the proposed algorithm is attributed to the employment of image local and nonlocal regularization at the same time, which offers a powerful mechanism of characterizing the statistical properties of natural images.

TABLE V. PSNR/FSIM Comparisons for Image Deblurring

| Image | Butterfly | Foreman | House | Leaves |
|---|---|---|---|---|
| 9×9 Uniform Kernel, $\sigma$ = 0.5 | | | | |
| SALSA [15] | 30.30/0.9300 | 33.21/0.9281 | 33.95/0.9398 | 29.02/0.9245 |
| SA-DCT [12] | 29.67/0.9261 | 34.07/0.9392 | 35.37/0.9335 | 28.99/0.9239 |
| BM3D [21] | 28.73/0.8959 | 34.18/0.9396 | 35.57/0.9363 | 29.32/0.9155 |
| Proposed | 31.03/0.9394 | 36.10/0.9612 | 37.73/0.9670 | 31.61/0.9520 |
| Gaussian Kernel: fspecial('Gaussian', 25, 1.6), $\sigma$ = 0.5 | | | | |
| SALSA [15] | 31.24/0.9541 | 32.31/0.9492 | 33.95/0.9386 | 30.32/0.9518 |
| SA-DCT [12] | 30.46/0.9373 | 32.63/0.9519 | 34.28/0.9288 | 30.50/0.9459 |
| BM3D [21] | 29.80/0.9106 | 32.91/0.9499 | 34.11/0.9302 | 30.61/0.9342 |
| Proposed | 31.26/0.9473 | 35.12/0.9664 | 36.68/0.9605 | 32.18/0.9610 |
| Motion Kernel: fspecial('motion', 20, 45), $\sigma$ = 0.5 | | | | |
| SALSA [15] | 30.97/0.9350 | 33.35/0.9335 | 33.58/0.9376 | 29.63/0.9312 |
| SA-DCT [12] | 30.75/0.9428 | 34.59/0.9484 | 35.17/0.9366 | 30.03/0.9384 |
| BM3D [21] | 29.71/0.9119 | 34.70/0.9491 | 34.81/0.9344 | 30.42/0.9316 |
| Proposed | 33.10/0.9572 | 37.28/0.9695 | 37.40/0.9668 | 33.95/0.9693 |

Furthermore, JSM model is compared with AKTV [46], which is known to work quite well in the case of large blur. Here, the case with 19×19 uniform PSF for image *Cameraman* is tested, with the corresponding BSNR equal to 40. BSNR means Blurred Signal to Noise Ratio, and is equivalent to 10*log (Blurred signal variance/Noise variance). Smaller BSNR means larger noise variance. The objective and visual quality comparisons are shown in Figs. 14. From Fig. 14, it is apparent to see that JSM model produces better results than AKTV with much sharper image edges and less annoying ringing artifacts.

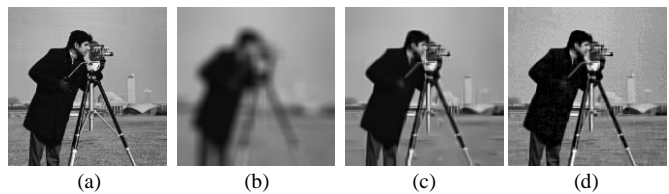

**Fig. 14.** Visual quality comparison of image deblurring on image *Cameraman* (19×19 uniform blur and BSNR=40). (a) Original; (b) Noisy and blurred; (c) AKTV (25.19 dB/0.8109) [46]; (d) Proposed (26.51 dB/0.8724).

### D. Mixed Gaussian plus Salt-and-Pepper Noise Removal

In practice, we often encounter the case where an image is corrupted by both Gaussian and salt-and-pepper noise. Such mixed noise could occur when an image that has already been contaminated by Gaussian noise in the procedure of image acquisition with faulty equipment suffers impulsive corruption during its transmission over noisy channels successively.

In our simulations, images will be corrupted by Gaussian noise with standard deviation $\sigma$ and salt-and-pepper noise density level $r$, where $\sigma$ is assumed to be known before and $r$ is unknown. For mixed Gaussian plus impulse noise, traditional image denoising methods that can only deal with one single type of noise don't work well due to the distinct characteristics of both types of degradation processes. Here, two state-of-the-art algorithms compared with our proposed method are: FTV [48] and IFASDA [49]. Experiments are carried out on four benchmark gray images in Fig. 7, where the standard variance σ of Gaussian noise equals 10 and the noise density level *r* varies from *40%* to *50%*. To handle this case, we first apply adaptive median filter [47] to the noisy image to identify the mask *H* , that is, change the problem of mixed Gaussian and impulse noise removal into the problem of image restoration from partial random samples with Gaussian noise, and then run the proposed algorithm according to Table II.

Table VI presents the PSNR/FSIM results of the three comparative denoising algorithms on all test images for Gaussian plus salt-and-pepper impulse noise removal. Obviously, the proposed method considerably outperforms the other methods in all the cases, with the highest PSNR and FSIM, achieving the average PSNR and FSIM improvements over the second best method (i.e., IFASDA) are 1.8 dB and 0.01, separately.

TABLE VI. PSNR/FSIM Comparisons for Gaussian plus Salt-and-Pepper Noise Removal

| *Image* | *Barbara* | *House* | *Boat* | *Lena* |
|---|---|---|---|---|
| *r*=40% and *σ* = 10 | | | | |
| **Noisy** | 9.36/0.4153 | 9.46/0.3499 | 9.42/0.5579 | 9.40/0.4848 |
| **FTV** [48] | 26.18/0.8899 | 31.10/0.9156 | 28.53/0.9405 | 30.85/0.9574 |
| **IFASDA** [49] | 28.59/0.9252 | 32.26/0.9263 | 30.28/0.9614 | 32.27/0.9625 |
| **Proposed** | **31.81/0.9481** | **34.33/0.9380** | **30.92/0.9635** | **33.49/0.9708** |
| *r*=50% and *σ* = 10 | | | | |
| **Noisy** | 8.39/0.3815 | 8.50/0.3170 | 8.46/0.5230 | 8.44/0.4501 |
| **FTV** [48] | 25.40/0.8728 | 30.36/0.9050 | 27.66/0.9259 | 30.20/0.9488 |
| **IFASDA** [49] | 27.45/0.9129 | 31.69/0.9181 | 29.50/0.9556 | 31.70/0.9579 |
| **Proposed** | **31.04/0.9383** | **33.72/0.9264** | **30.12/0.9613** | **32.90/0.9645** |

Some visual results of the recovered images for the three algorithms are presented in Fig. 15. One can see that FTV [48] is effective in suppressing the noises; however, it produces over-smoothed results and eliminates much image details (see Fig. 15(b)). IFASDA [49] is very competitive in recovering the image structures. However, it tends to generate some annoying artifacts in the smooth regions (see Fig. 15(c)). By comparing with TV and IFASDA, the proposed method provides the most visually pleasant results (see Fig. 15(d)).

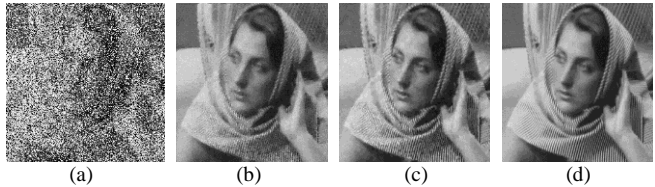

(a)　　　　(b)　　　　(c)　　　　(d)

**Fig. 15.** Visual quality comparison of mixed Gaussian plus salt-and-peppers impulse noise removal on image *Barbara*. (a) Noisy image corrupted by Gaussian plus salt-and-pepper impulse noise with *σ* = 10 and *r* = *50%*; (b)–(d) Denoised results by FTV (25.40 dB/0.8728) [48], IFASDA (27.45 dB/0.9129) [49], and the proposed algorithm (31.04 dB/0.9383).

### E. Parameter Optimization

In our proposed algorithm, we have four parameters to determine, i.e., $\tau$, $\lambda$, $\mu_1$ and $\mu_2$, which seems quite complicated. To make it tractable, we simplify the optimization of four parameters into the optimization of one parameter $\tilde{\mu}$. Specifically, in Eq. (16), to make a tradeoff between LSM and NLSM,

$\mu_1$ and $\mu_2$ in the ratio of one to six is exploited, which is verified by our experiments. Moreover, due to the relationship $\tilde{\mu} = \mu_1 + \mu_2$, we get $\mu_1 = 0.14\tilde{\mu}$ and $\mu_2 = 0.86\tilde{\mu}$. To determine $\tau$ and $\lambda$, we observe that, the standard deviation $\sigma$ of Gaussian noise $n$ in Eq. (1) is not larger than ten, a good rule of thumb is $\tau = 10\mu_1$, $\lambda = 10\mu_2$ [15]. Therefore, it yields $\tau = 1.4\tilde{\mu}$ and $\lambda = 8.6\tilde{\mu}$. So far, the relationships between the above four parameters and $\tilde{\mu}$ are established. In practice, for each case of image processing application, the optimization of $\tilde{\mu}$ is obtained by simply searching some values.

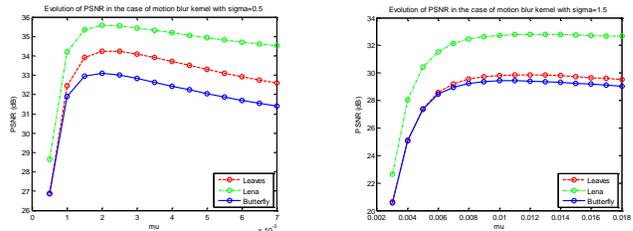

**Fig. 16.** PSNR evolution with respect to parameter $\tilde{\mu}$ in the cases of motion blur kernel with Gaussian noise standard deviation *σ* = 0.5 and *σ* = 1.5 for three test images.

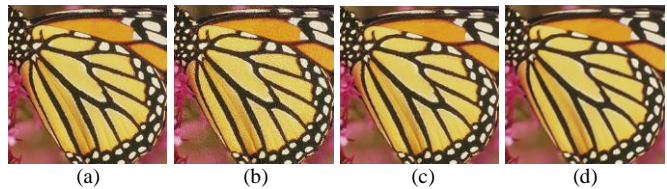

(a)　　　　(b)　　　　(c)　　　　(d)

**Fig. 17.** Visual quality comparison of proposed algorithm with various $\tilde{\mu}$ in the case of image deblurring with motion blur kernel and *σ* = 0.5. (a) Original image; (b) Deblurred result with $\tilde{\mu}$ =5e-4, PSNR=26.87; (c) Deblurred result with $\tilde{\mu}$ =2e-3, PSNR=33.10; (d) Deblurred result with $\tilde{\mu}$ =3e-2, PSNR=28.50.

Take the case of image deblurring as example. Fig. 16 provides PSNR evolution with respect to $\tilde{\mu}$ in the cases of motion blur kernel with Gaussian noise standard deviation *σ* = 0.5 and *σ* = 1.5 for three test images. From Fig. 16, three conclusions can be observed. First, as expected, there is an optimal $\tilde{\mu}$ that achieves the highest PSNR by balancing image noise suppression with image details preservation (see Fig. 17(c)). That means, if $\tilde{\mu}$ is set too small, the image noise can't be suppressed (see Fig. 17 (b)); if $\tilde{\mu}$ is set too large, the image details will be lost (see Fig. 17(d)). Second, in each case, the optimal $\tilde{\mu}$ for each test image is almost the same. For instance, in the case of *σ* = 0.5, the optimal $\tilde{\mu}$ is 2e-3, and in the case of *σ* = 1.5, the optimal $\tilde{\mu}$ is 1e-2. This is very important for parameter optimization, since the optimal $\tilde{\mu}$ in each case can be determined by only one test image and then applied to other test images. Third, it is obvious to see that $\tilde{\mu}$ has a great relationship with σ. A larger σ corresponds to a larger $\tilde{\mu}$.

### F. Algorithm Complexity and Computational Time

Comparing the $u$, $w$, $x$ sub-problems, it is obvious to conclude that the main complexity of the proposed algorithm comes from the $x$ sub-problem, which requires the operations of 3D transforms and inverse 3D transforms for each 3D array. In our implementation, for image *House* with size 256×256, each iteration costs about 1.25 s on a computer with Intel

3.25GHz CPU. Take image inpainting application for example. With degraded images as default initialization described by Table VII, it takes about 130 s by 100 iterations in the case of *Ratio=80%* and about 510 s by 400 iterations in the case of *Ratio=20%*. All the computational time for image *House* with various methods are given in Table VII.

TABLE VII. Computational Time Comparisons of Different Methods (unit: s)

| Image | Ratio | MCA | SKR | BPFA | FoE | Degraded +JSM | SKR +JSM |
|---|---|---|---|---|---|---|---|
| House | 20% | 237.8 | 10.8 | 2170.4 | 217.4 | 506.8 | 75.1 |
| | 30% | 224.1 | 11.2 | 2200.6 | 220.6 | 443.5 | 49.2 |
| | 50% | 209.4 | 12.1 | 2260.5 | 226.5 | 316.7 | 37.4 |
| | 80% | 196.6 | 13.4 | 2280.9 | 228.9 | 126.7 | 26.1 |
| Avg. | | 216.9 | 11.8 | 2240.4 | 224.4 | 348.4 | 46.9 |

To speed up our proposed algorithm, on one hand, we can exploit the results of SKR instead of degraded images as initialization, which decreases the number of iteration enormously. The last column of Table VII shows the computational time, which is about one seventh of the original time (denoted by the column next to the last). On the other hand, ongoing work addresses the parallelization, utilizing GPU hardware to accelerate the proposed algorithm.

*G. Algorithm Convergence and Robustness*

From the discussions above, the computational time of the proposed algorithm would be significantly reduced along with a good initialization. In this sub-section, we will verify the convergence and robustness of the proposed algorithm.

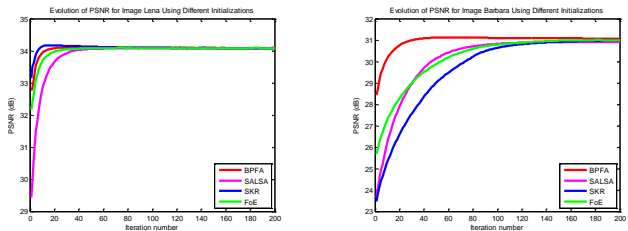

**Fig. 18.** Verification of the convergence and robustness of the proposed algorithm. From left to right: progression of the PSNR (dB) results achieved by proposed algorithm with various initializations with respect to the iteration number in the cases of image inpainting with ratio=0.3 for images *Lena* and *Barbara*.

Take the cases of image inpainting application when *Ratio=30%* for two images *Lena* and *Barbara* as examples. The restoration results generated by SALSA [15], FoE [36], SKR [35], BPFA [38] are utilized as initialization for the proposed algorithm, respectively. Fig. 18 plots the evolutions of PSNR versus iteration numbers for test images with various initializations. It is observed that with the growth of iteration number, all the PSNR curves increase monotonically and almost converge to the same point, which fully demonstrates the convergence of the proposed algorithm. The algorithm convergence also makes the termination of the proposed algorithm easier, which just needs to reach the preset maximum iteration number. Furthermore, it is obvious that the initialization results with higher quality require fewer iteration numbers to be convergent. The tests fully illustrate the robustness of our proposed method, that it, our proposed method is able to provide almost the same results when starting with various initializations.

V. CONCLUSIONS

In this paper, a novel algorithm for high-quality image restoration using joint statistical modeling in space-transform domain is proposed, which efficiently characterizes the intrinsic properties of local smoothness and nonlocal self-similarity of natural images from the perspective of statistics at the same time. Experimental results on three applications: image inpainting, image deblurring and mixed Gaussian and salt-and-pepper noise removal have shown that the proposed algorithm achieves significant performance improvements over the current state-of-the-art schemes and exhibits nice convergence property. Future work includes the investigation of the statistics for natural images at multiple scales and orientations and the extensions on a variety of applications, such as image deblurring with mixed Gaussian and impulse noise and video restoration tasks.

ACKNOWLEDGMENT

The authors would like to thank the authors of [12], [15], [21], [28], [35], [36], [37], [38], [46], and [52] for kindly providing their codes. The authors would also like to thank the anonymous reviewers for their helpful comments and suggestions.